\documentclass[aps,prl,superscriptaddress,reprint]{revtex4-1}
\usepackage{amsmath}
\usepackage{amssymb}
\usepackage{graphicx}
\usepackage{hyperref}
\usepackage{url}
\usepackage{color,xcolor}
\usepackage{float}
\usepackage{ulem}

\makeatletter
\newcommand{\figcaption}{\def\@captype{figure}\caption}
\newcommand{\tabcaption}{\def\@captype{table}\caption}
\makeatother

\begin{document}

\title{Frustrated Dipole Order Induces Noncollinear Proper Ferrielectricity in Two Dimensions}
\author{Ling-Fang Lin}
\author{Yang Zhang}
\affiliation{School of Physics, Southeast University, Nanjing 211189, China}
\affiliation{Department of Physics and Astronomy, University of Tennessee, Knoxville, Tennessee 37996, USA}
\author{Adriana Moreo}
\author{Elbio Dagotto}
\affiliation{Department of Physics and Astronomy, University of Tennessee, Knoxville, Tennessee 37996, USA}
\affiliation{Materials Science and Technology Division, Oak Ridge National Laboratory, Oak Ridge, Tennessee 37831, USA}
\author{Shuai Dong}
\email{Corresponding author. Email: sdong@seu.edu.cn}
\affiliation{School of Physics, Southeast University, Nanjing 211189, China}
\date{\today}

\begin{abstract}
Within Landau theory, magnetism and polarity are homotopic, displaying a one-to-one correspondence between most physical characteristics. However, despite widely reported noncollinear magnetism, spontaneous noncollinear electric dipole order as ground state is rare. Here a dioxydihalides family is predicted to display noncollinear ferrielectricity, induced by competing ferroelectric and antiferroelectric soft modes. This intrinsic noncollinearity of dipoles generates unique physical properties, such as $\mathbb{Z}_2\times\mathbb{Z}_2$ topological domains, atomic-scale dipole vortices, and negative piezoelectricity.
\end{abstract}

\maketitle

\begin{figure}
\centering
\includegraphics[width=0.47\textwidth]{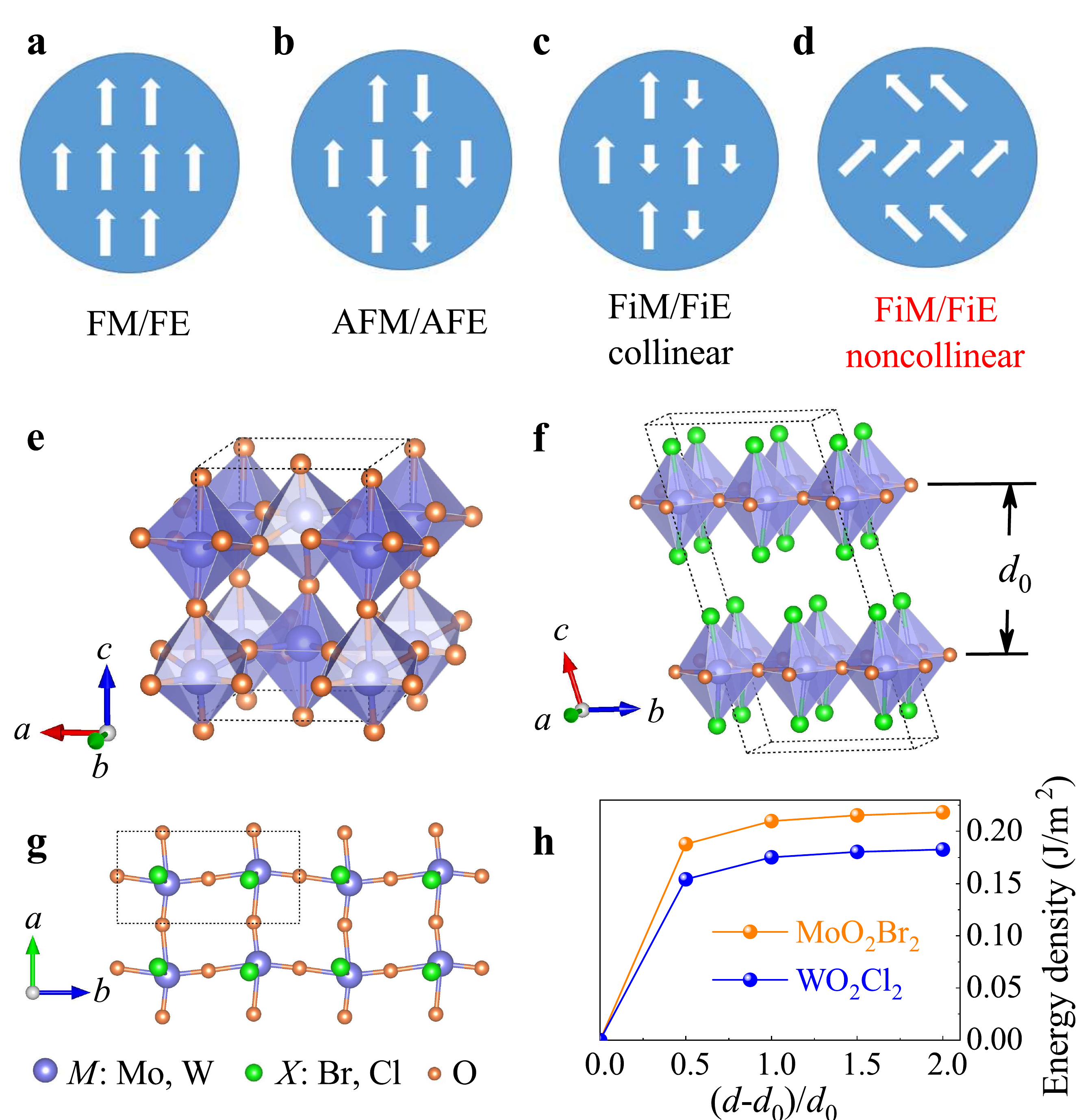}
\caption{(a-d) Sketch of the spin/dipole orders discussed in the text.
(a) FM/FE parallel alignment; (b) AFM/AFE antiparallel alignment,
fully compensated; (c-d) FiM/FiE: similar to (b) but with
magnetization/polarization only partially compensated.
(e, f) Structures of $M$O$_nX_{6-2n}$. (e) $n=3$; (f) $n=2$. The most stable vdW
stacking is the A-B type. (g) Top view of a dioxydihalide monolayer
(dash lines indicate a unit cell). (h) Cleavage energy density of the $\alpha$ phase.}
\label{structure}
\end{figure}

The Landau theory of phase transitions provides an elegant common framework for both magnetic and polar systems. The one-to-one correspondence between physical characteristics, such as ordered phases -- ferromagnetic (FM) {\it vs} ferroelectric (FE) states, antiferromagnetic (AFM) {\it vs} antiferroelectric (AFE) states [see Figs.~\ref{structure}(a-b)]--, hysteresis loops, domains, and other properties is well recognized. However, ferrielectric (FiE) systems, with partially compensated collinear dipoles [Figs.~\ref{structure}(c-d)], are rare (except in liquid crystals
and in a few solids like hybrid improper ferroelectrics~\cite{Benedek:Prl})~\cite{Scott:Prb},
while ferrimagnetic (FiM) materials are fairly common, e.g. Fe$_3$O$_4$.

This incompleteness of dipole orders is even more dramatic with regards to {\it noncollinearity}. For magnets, spin noncollinearity has been widely studied \mbox{\cite{Ramirez:Arms,Cheong:Nm}}, leading to exotic magnetism-driven polarization ($P$) \cite{Kimura:Armr}, skyrmions \cite{Nagaosa:Nn}, and topological anomalous Hall effect \cite{Nagaosa:Rmp}. There are several mechanisms to generate these crucial noncollinear spin orders. For example, in geometrically frustrated systems, such as two-dimensional (2D) triangular lattices, the AFM coupling between nearest-neighbor (NN) spins can generate the $120^\circ$ order \cite{Ramirez:Arms}. For other lattices, the exchange frustration, typically involving competition between NN FM ($J_1$) and next-nearest-neighbor (NNN) AFM ($J_2$) couplings, can generate magnetic cycloid or helical arrangements \cite{Cheong:Nm}.

By contrast, the electric dipoles within a FE or AFE domain always tend to be
parallel or antiparallel, aligned by the dipole-dipole
interactions \cite{Dawber:Rmp,Rabe:Bok}. Although slightly noncollinear dipole
orders were proposed in a few FiEs, e.g. BaFe$_2$Se$_3$ and
Pb$_2$MnWO$_6$ \cite{Dong:PRL14,Orlandi:Ic}, their noncollinearities
are rigidly fixed by the local crystalline environment and, thus, can be
trivially modulated. Noncollinear FiE phases were also predicted for strained
BiFeO$_3$ \cite{Yangyurong:Prl,Yangyurong14:Prl,Prosandeev:Prl}, which has attracted much interest while waiting for experimental verification. In addition, dipoles can become noncollinear at some domain walls, as when forming flux-closure domains and even dipole vortices/skymions \cite{Jia:Science,Naumov:Nature,Nahas:Nc,Yadav:Nature,Das:Nature}.
However, such noncollinearity is not a primary property of the FE state
but driven by electrostatic effects from geometrically-confined boundary conditions.

Inspired by the ``frustration" concept from magnets, here a series of 2D materials ($M$O$_2X_2$,
$M$: group-VI transition metal; $X$: halogen) is studied theoretically, which we predict can host intrinsic noncollinear electric dipole textures {\it spontaneously}.

2D FE monolayers (or few-layers) exfoliated from van der Waals (vdW) layered materials are intrinsically superior at the nanoscale as compared with canonical three-dimensional (3D) FE materials \cite{Chang:Science,Liu:Nc,Ding:Nc,Zhou:Nl,Wu:Wcms}. In spite of this potential value, 2D FE materials remain rare \cite{Wu:Wcms} and their physical properties,
such as domain structures, have not been well explored. This Letter demonstrates that these particular 2D noncollinear polar systems
can provide an ideal platform to explore exotic polarity and topological domains beyond the standard collinear ferroelectricity,
which may be crucial for domain wall nanoelectronics \cite{Catalan:Rmp}.

\textit{Physical properties.-}
 Starting from the 3D $M$O$_3$ crystal [perovskite-like structure without A-site ions, see Fig.~\ref{structure}(e)], $M$O$_2X_2$ can be derived by replacing the apical O$^{2-}$ oxygens by double halide ions $X^-$ [Fig.~\ref{structure}(f)], forming a series of vdW layered materials. Further replacement of O$^{2-}$ by $X^-$ can lead to quasi-one-dimensional $M$O$X_4$ chains and to the zero-dimensional molecular limit $MX_6$ [see Figs.~S1(a-b) in Supplementary Material (SM)~\cite{Supp}].

\begin{figure*}
\centering
\includegraphics[width=\textwidth]{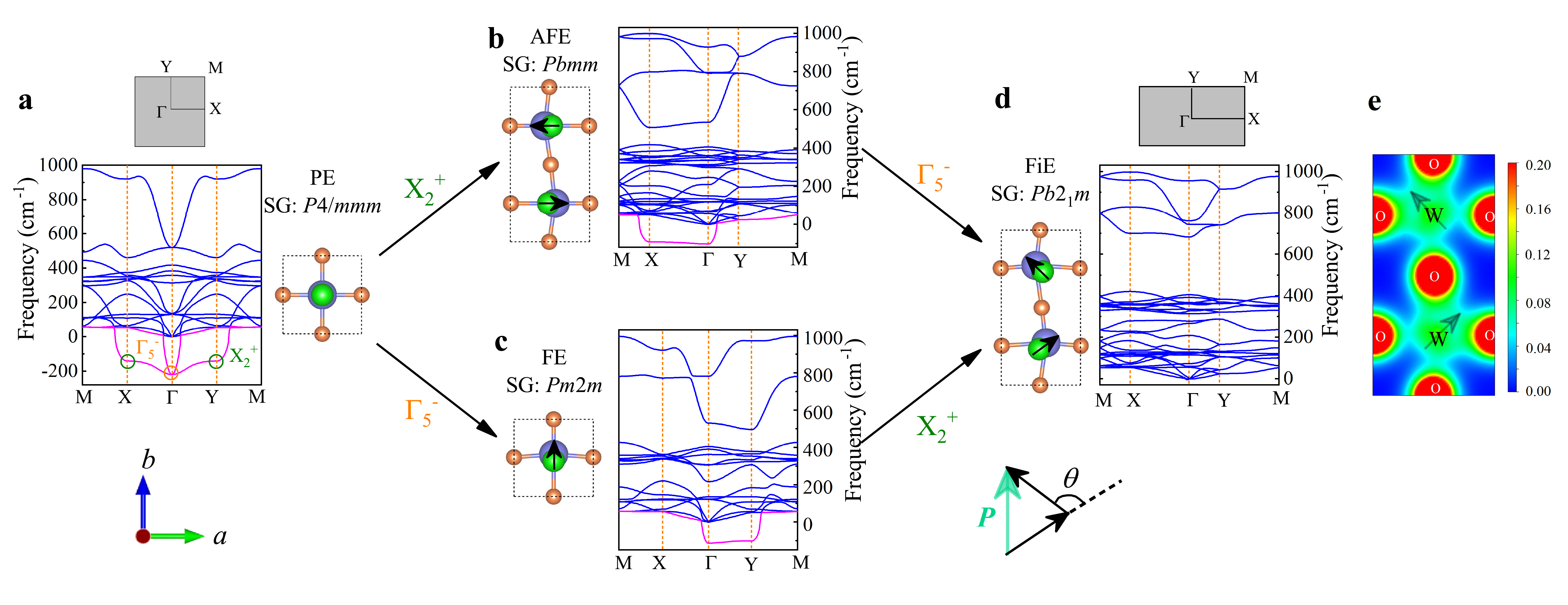}
\caption{Structural transitions of a $M$O$_2X_2$ monolayer. Shown are the phonon spectra
of WO$_2$Cl$_2$ (those of MoO$_2$Br$_2$ are very similar, see Fig.~S4 in SM \cite{Supp}).
The imaginary-frequency branches of phonons are displayed in pink. Grey square/rectangles
are Brillouin zones. (a) The PE phase. The unstable distortion modes ($X_2^+$ \& $\Gamma_5^-$) are indicated. (b) The intermediate AFE phase with the $X_2^+$ distortion mode. (c) The intermediate FE phase with the $\Gamma_5^-$ distortion mode. (d) The resulting stable FiE ground state with both the $X_2^+$ \& $\Gamma_5^-$ distortion modes. The angles $\theta$ between NN dipoles (black arrows) along the $b$-axis are $\sim129^\circ$. (e) Top view of the valence charge density of the WO$_2$ plane  (integrated from $-8$ eV to the Fermi level). The $d^0$ orbitals of W$^{6+}$ tend to form coordination bonds with two neighboring oxygens.}
\label{phonon}
\end{figure*}

In this dioxydihalides family, WO$_2$Cl$_2$ and MoO$_2$Br$_2$ have been synthesized
experimentally~\cite{Jarchow:Zaac,Armstrong:Aciee,Schustereit:Zaac}, while the crystal
structures of the corresponding fluorides, iodides, and CrO$_2X_2$ are isomeric or
unknown. Thus, in the following only WO$_2$Cl$_2$ and MoO$_2$Br$_2$ will be studied
using density functional theory (DFT)~\cite{Supp}. Most previous studies of $M$O$_2X_2$ were devoted only to their chemical properties,
while their physical properties were rarely addressed.

As shown in Fig.~\ref{structure}(g), each $M$O$_2X_2$ layer is composed of
corner-sharing octahedra~\cite{Schustereit:Zaac}. Contrary to most 2D materials
which display compact honeycomb or triangular atomic arrangements, the ``square''
lattice of $M$O$_2X_2$ is spatially loose, which is advantageous for polar distortions.
Due to the weak vdW interactions, there are two stacking modes for $M$O$_2X_2$ layers
observed in experiments: the A-B stacking $\alpha$-type (Fig.~\ref{structure}(f)] and
the A-A stacking $\beta$-type [Fig.~S1 in SM \cite{Supp}), corresponding to the space groups $Bb$ (No. $9$) and $Pb2_1m$ (No. $26$), respectively~\cite{Schustereit:Zaac,Jarchow:Zaac}. The optimal distances ($d_0$'s) between adjacent layers are shorter in the
$\alpha$ phases (see Table~S1 in SM \cite{Supp}), which are lower in energy
than the $\beta$ phases by $70$ meV/W and $77$ meV/Mo.

The cleavage energies were calculated to analyze whether it is possible
to exfoliate $M$O$_2X_2$ monolayers, as shown in Fig.~\ref{structure}(h).
For the $\alpha$ phase, the cleavage energies are $0.22$ J/m$^2$
and $0.18$ J/m$^2$  for MoO$_2$Br$_2$ and WO$_2$Cl$_2$, respectively.
For comparison, the cleavage energy for graphite is $0.325$ J/m$^2$
theoretically~\cite{Mounet:Nn} and $0.37$ J/m$^2$
experimentally~\cite{Wang:Nc}. Thus, the exfoliation of a monolayer, or few-layers,
from bulk $M$O$_2X_2$ should be feasible experimentally. Furthermore, our molecular
dynamic simulation confirms the thermal stabilities of these monolayers at $300$~K and $400$ K (Fig.~S2 in SM \cite{Supp}).

Additional physical properties of $M$O$_2X_2$ monolayers and bulk forms are summarized in SM (Fig.~S3 and Table. S2) \cite{Supp}.

\textit{Noncollinear dipole order.-}
The paraelectric (PE) structure of $M$O$_2X_2$ is shown
in Fig.~\ref{phonon}(a), where all $M$ ions are restored to the central positions
of the O$_4X_2$ octahedra. In its phonon spectrum there are two imaginary-frequency
branches, which will lead to spontaneous distortions. The symmetric $B_{1g}$ (i.e. $X_2^+$)
phonon mode at $X$ (and $Y$) leads to AFE-type distortions [Fig.~\ref{phonon}(b)]. The
double-degenerate $E_u$ (i.e. $\Gamma_5^-$) phonon mode at $\Gamma$ leads to FE-type
distortions [Fig.~\ref{phonon}(c)]. Both these distortions lower the symmetry from
tetragonal to orthorhombic.

The most striking physical result is that the cooperation of these two distortion modes,
which resemble the exchange frustration in magnets, leads to a net FiE structure
[Fig.~\ref{phonon}(d)], which is dynamically stable according to its phonon spectrum.
In this FiE state, the AFE and FE ordering directions are orthogonal, along the $a$ and
$b$ axes, respectively. If the vdW $d_0$ is used as the thickness of a monolayer, the
calculated net $P$'s along the $b$ axis are slightly larger than their bulk values
(see Table~S2 in SM \cite{Supp}), themselves only a little larger than those of
BaTiO$_3$ ($\sim 20-25$ $\mu$C/cm$^2$).

The $d^0$ rule (i.e. the formation of coordination bonds between $M$'s empty $d$
orbitals and O's $2p$ orbitals) should be the
driving force for the polar distortions. This was confirmed by the Bader charge
calculation~\cite{Henkelman:Cms}.
As shown in Table~S3 in SM \cite{Supp}, a small portion of valence electrons ``leak''
from O$^{2-}$ to $M^{6+}$, accompanying the FiE
distortions, which can also be visualized in Fig.~\ref{phonon}(e).
Although the $d^0$ rule is well known for FE/AFE perovskites,
such as BaTiO$_3$ and PbZrO$_3$, interestingly none of the previously studied
2D FE systems belongs to this $d^0$ category \cite{Wu:Wcms},
since none have perovskite-like structures.

The novel FiE state unveiled here is exotic. Due to the frustration between FE and AFE modes,
here the local dipole moments are {\it noncollinear}, with a canting angle $\sim129^\circ$
for WO$_2$Cl$_2$ ($\sim128^\circ$ for MoO$_2$Br$_2$) between NN dipoles along the $b$ axis.
This unexpected noncollinearity leads to unique physics, as discussed below.

\begin{figure}
\centering
\includegraphics[width=0.47\textwidth]{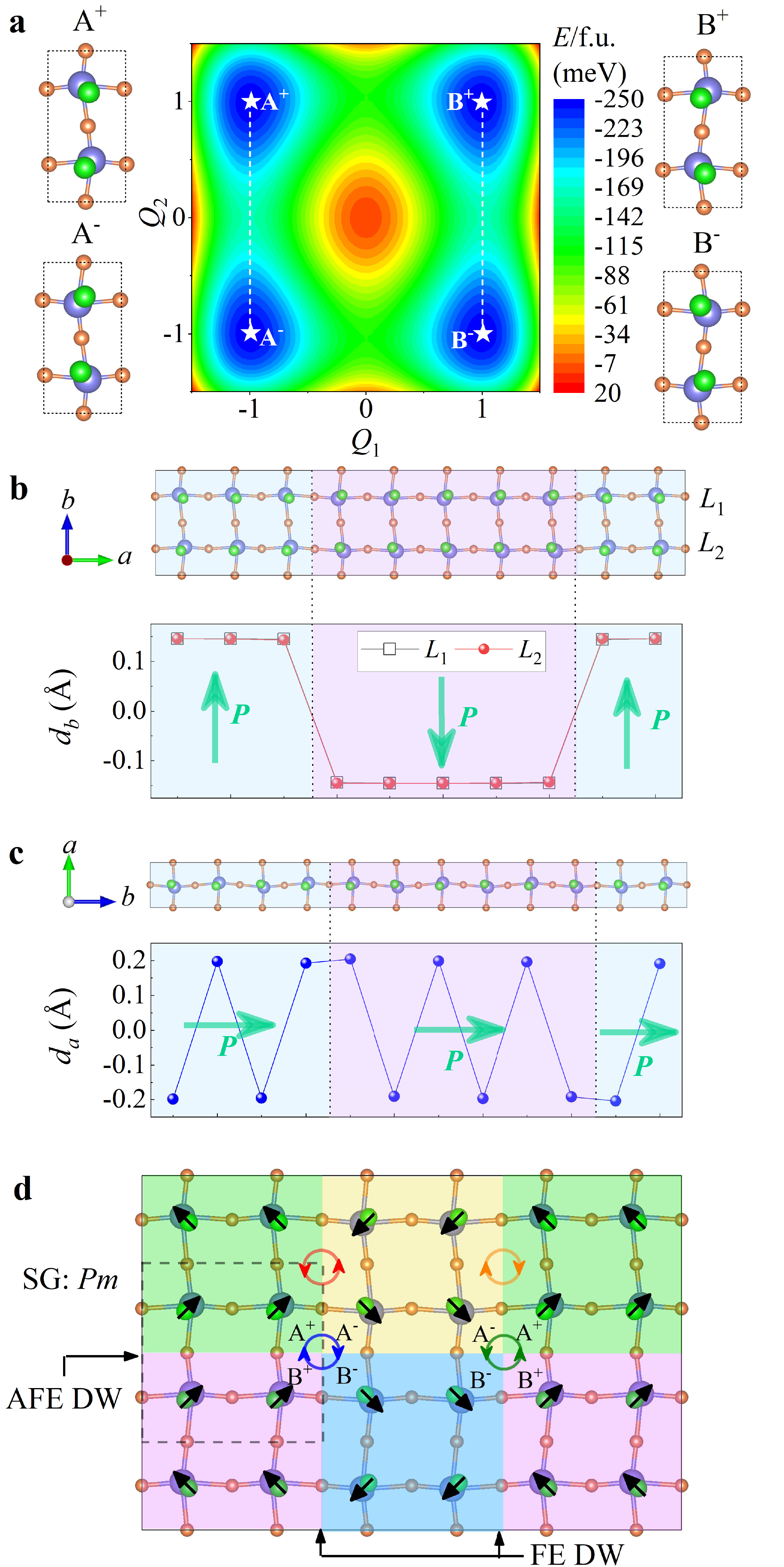}
\caption{Formation of $\mathbb{Z}_2\times\mathbb{Z}_2$ domain textures in WO$_2$Cl$_2$.
Results in MoO$_2$Br$_2$ are qualitatively similar (Fig.~S5 in SM \cite{Supp}). (a) Contour
plot of the energy varying $Q_1$ and $Q_2$ (amplitudes of $X_2^+$ and $\Gamma_5^-$ modes).
$Q_1$ and $Q_2$ are normalized to their optimized values. The indicated four
degenerate lowest-energy wells correspond to the four domains (A$^+$, A$^-$, B$^+$, and B$^-$).
$+$/$-$ indicate the sign of $P$ along the $+b$ axis, while $A$/$B$ distinguish the AFE
configuration. (b-c) The atomically sharp domain walls: (b) FE domain walls; (c) AFE domain
walls [$d_a$ ($d_b$) are the displacements along the $a$-axis ($b$-axis)]. (d) Schematic of
four-colored domains. The FE and AFE domain walls form perpendicular crossovers. Red (orange)
circles denote dipoles that form a vortex (antivortex) at the FE domain walls. The broken line
square contain FE double-stripes at AFE domain walls. The blue (green) circle is a topological
$\mathbb{Z}_2\times\mathbb{Z}_2$ antiphase domain emerging from one core. Two kinds of cores (vortex \& antivortex) can be defined by the chirality of the domain phase.}
\label{domain}
\end{figure}

\textit{Domain \& domain walls.-}
The DFT energy contour as a function of distortion modes for WO$_2$Cl$_2$ is shown
in Fig.~\ref{domain}(a) and its Landau energy
fitting can be found in SM \cite{Supp}. For selected orthorhombic axes, four degenerate
wells exist (characterized by the $\pm Q_1$ and $\pm Q_2$ of the $X_2^+$ and $\Gamma_5^-$ modes), corresponding to the four domains: A$^+$, A$^-$, B$^+$, and B$^-$. An intuitive conclusion arising from Fig.~\ref{domain}(a) is that the favoured domain walls are between A$^+$/A$^-$, A$^+$/B$^+$, B$^+$/B$^-$, and A$^-$/B$^-$, while domain walls between A$^+$/B$^-$ or A$^-$/B$^+$ are highly energetic,
thus unfavourable. A similar situation occurs in hexagonal
manganites~\cite{Choi:Nm,Artyukhin:Nm}, whose energy contour shows a Mexican-hat
sixfold symmetry and, thus, prefers the well-known $\mathbb{Z}_2\times\mathbb{Z}_3$
topological domain patterns \cite{Choi:Nm,Artyukhin:Nm}. Contrary to WO$_2$Cl$_2$,
the collaborative modes in hexagonal manganites are the FE distortion and trimerization
of the Mn-sublattice, not the AFE mode.

The possible FE and AFE domain walls in WO$_2$Cl$_2$ are shown in Fig.~\ref{domain}(b-c). Because the head-to-head and tail-to-tail charged domain walls are highly energetically unfavorable, we consider only the shoulder-by-shoulder charge-neutral domain walls. More details can be found in \cite{Supp}. According to the DFT structural relaxations, these domain walls are atomically sharp, leading to distinct domains. The domain wall energies for FE and AFE domain walls are $1.6$ meV/bond and $5.0$ meV/bond for WO$_2$Cl$_2$ ($11.5$ meV/bond and $8.5$ meV/bond for MoO$_2$Br$_2$), respectively.

As a consequence, it is natural to expect $\mathbb{Z}_2\times\mathbb{Z}_2$ topological domain patterns in $M$O$_2X_2$, as sketched in Fig.~\ref{domain}(d). If crystalline twinning and high-energetic charged domains can be excluded, the domain structure will be quite regular: the FE domain walls can only propagate along the $b$ axis while the AFE domain walls can only propagate along the $a$ axis, forming perpendicular cross points, namely atomic-scale $\mathbb{Z}_4$ antiphase
vortices/antivortices. More details of domains and domain walls can be found in the SM~\cite{Supp}.

In addition, atomic-scale dipole vortices/antivortices form at the FE domain walls, which is different to the dipole vortices
in PbTiO$_3$/SrTiO$_3$ superlattices and BiTiO$_3$ nanostructures,  whose
 size scale is much larger ($\sim 3-5$~nm) \cite{Naumov:Nature,Nahas:Nc,Yadav:Nature,Das:Nature}.
Our dipole vortices/antivortices are also different from those in the predicted $P2_12_12_1$ phase of strained BiFeO$_3$, whose atomic-scale
dipole vortices/antivortices form a closely packed array with a fixed position in the whole crystal \cite{Prosandeev:Prl},
while ours exist only at some domain walls and thus are movable. The study of a dipole vortex, in correspondence to a magnetic vortex (or skyrmion) in magnetism, is an emerging topic in the field of polar materials. FiE $M$O$_2X_2$ can provide a superior playground for the dipole vortex due to its intrinsic non-collinearity.

\begin{figure}
\centering
\includegraphics[width=0.47\textwidth]{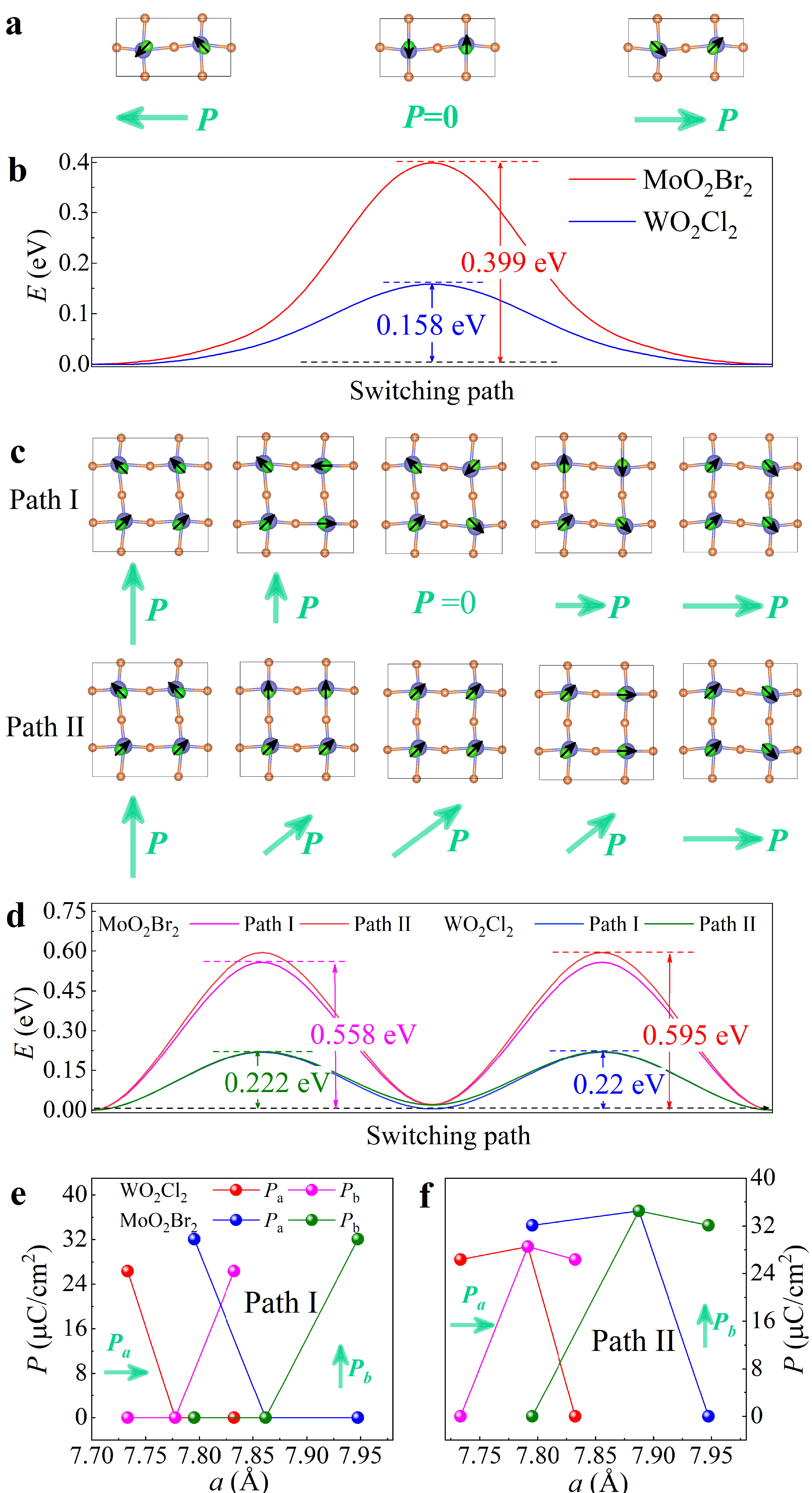}
\caption{Illustration of the unique FiE properties of $M$O$_2X_2$. (a) $180^\circ$ reversal path of the net $P$: FiE-AFE-FiE.
(b) Energy barriers of the $180^\circ$ reversal of $P$ (for two $M$'s). (c) Two possible two-step paths for the $90^\circ$ flop
of $P$, namely the interchange of crystalline axes $a$ and $b$. (d) Energy barriers for the $90^\circ$ flop of $P$ (for four $M$'s).
(e-f) Correspondence between the $a$/$b$ components of the $P$ vector and the lattice constant $a$. The overall tendency is that $P_a$
decreases with increasing $a$ -- negative piezoelectricity -- although the actual process is unknown. (e) Path~I. (f) Path~II.}
\label{switch}
\end{figure}

\textit{Negative piezoelectricity.-}
For applications of FE materials, the switching of $P$'s are important physical properties.
Interestingly, the switching paths for $M$O$_2X_2$ are nontrivial.
As sketched in Fig.~\ref{switch}(a), the $180^\circ$ reversal of the net $P$
does not require the $180^\circ$ flip of local dipoles, different
from the plain FE cases. Instead, the collaborative $\sim50^\circ$
rotations of local dipoles are enough,
another unique property of the unveiled noncollinear ferrielectricity.
As shown in Fig.~\ref{switch}(b), the AFE state can act as the nonpolar intermediate state,
leading to a moderate energy barrier for $P$ reversal.
Comparing with their bulk values, the energy barriers
do not change much, namely from $77$ meV/f.u. (bulk) to $79$ meV/f.u. (monolayer)
for WO$_2$Cl$_2$.

Another interesting issue is the $90^\circ$ switching of $P$, which corresponds
to the interchange of crystalline axes $a$ and $b$. Two possible
two-step paths are proposed to achieve this flop. As shown in Fig.~\ref{switch}(c),
in each step half of the dipoles rotate while the other half
remain fixed. Interestingly, the middle states just correspond
to two kinds of domain walls, as sketched in Fig.~\ref{domain}(d).
The corresponding switching energy barriers are plotted in Fig.~\ref{switch}(d),
suggesting a slightly lower barrier for path I.

An interesting property of this $90^\circ$ flop of $P$ is the resulting negative
piezoelectricity due to noncollinear ferrielectricity.
Starting from the tetragonal PE state, the elongated axis of the orthorhombic
FiE state is perpendicular to the direction of net $P$,
i.e. $a>b$ for $P||b$, which is unusual as compared with most ferroelectrics.
This anomalous behavior is due to the noncollinearity of
local dipoles, since the $a$-axis component of each local dipole is larger
than the $b$-axis component. Then, the $90^\circ$ flop of
$P$ will lead to a negative piezoelectric coefficient $d_{33}$
(see Figs.~\ref{switch}(e-f)), at least for
the partial intermediate process during the $90^\circ$ flop. In recent years,
the existence of exotic negative piezoelectricity was
predicted in special materials~\cite{Liu:Prl}, but experimentally
has only been observed in the organic FE polymer
poly(vinylidene fluoride) (PVDF)~\cite{Katsouras:Nm} and CuInP$_2$S$_6$~\cite{You:Sa}.

Our calculations predict that the monolayer dioxydihalides $M$O$_2X_2$'s are promising 2D polar materials with exotic noncollinear ferrielectricity. The $d^0$ rule, which works in 3D ferroelectric perovskites but has not been found in 2D ferroelectrics before, is the driving force for the polar distortions. Noncollinear ferrielectric order persisting up to room temperature are expected. More importantly, the frustration between the FE and AFE modes generates an {\it intrinsically noncollinear dipole texture}, which leads to unique physics in the novel FiE state unveiled here, such as $\mathbb{Z}_2\times\mathbb{Z}_2$ antiphase domain vortices, atomic-level dipole vortices at domain walls, negative piezoelectricity, and others. The key idea introduced here --the frustration of phonon instabilities -- provide one more route to pursuit a variety of noncollinear dipole orders, such as cycloid dipole textures with particular chiralities or even dipole-based skyrmions. As a consequence of this noncollinearity, exotic physics is expected to emerge.

\begin{acknowledgments}
We thank Prof. Y. G. Yao and Dr. S. Guan for illuminating discussions. This work was primarily supported by the National Natural Science Foundation of China (Grant Nos. 11834002 and 11674055). A.M. and E.D. were supported by the U.S. Department of Energy (DOE), Office of Science, Basic Energy Sciences (BES), Materials Science and Engineering Division. L.F.L. and Y.Z. were also supported by the China Scholarship Council. We thank the Tianhe-II of the National Supercomputer Center in Guangzhou (NSCC-GZ)
and the Big Data Center of Southeast University for providing the facility support on the numerical calculations.
\end{acknowledgments}

\bibliographystyle{apsrev4-1}
\bibliography{ref3}
\end{document}